\documentclass[twocolumn,fleqn]{article}

\begin{document}
\twocolumn[
{\Large\bf
Snyder-Yang algebra and confinement of color particles     
}

\bigskip

{\large V.V. Khruschov$^{a, b}$}

\medskip

{\it a- National Research Centre "Kurchatov Institute",
1 Kurchatov Sq., 123182 Moscow, Russia

b- Centre for Gravitation and Fundamental
Metrology, VNIIMS, 46 Ozyorn.St.,
119361 Moscow, Russia
}

\medskip

\vspace*{-0.5cm}

\begin{center}
\begin{abstract}
\parbox{15cm}{A model of color particle confinement is considered.
The model is based on the Snyder-Yang algebra, which takes into account
a non-commutativity of generalized momenta and coordinates of a color 
particle and contains two new constants. An extended kinematical 
invariance in a quantum phase space of a color particle give rise 
to an invariant equation with an oscillator rising potential. 
The presence of the oscillator rising potential can simulate 
a confinement of a color particle. Mass and lenght parameters involved 
in the Snyder-Yang commutation relations along with parameters of  
current and constituent quarks are estimated.



}

\end{abstract}
\end{center}
\bigskip
\smallskip
]

\smallskip
\begin{center}
\noindent {\bf 1. INTRODUCTION}
\end{center}
\smallskip

An origin of confinement as  interesting physical phenomenon is under
active study  from the begining of the QCD era \cite{alko}. Today we have no 
rigorous  proof  of this fact in spite of considerable efforts.
Color particle confinement is investigated
in the frame of different approaches, for instance, such as
the lattice QCD, the Schwinger-Dyson equations, 
massive transverse gluons, potential models, {\it etc}. 
In the present paper we propose a model for 
color particle confinement based on the Snyder-Yang algebra 
(SYA) \cite{sny,yang}
for color particle operators which have the meaning of angular momenta and
generalized momenta and coordinates.

The SYA maintains the non-commutativity of generalized momenta and coordinates
and requires  two new constants with dimensionality of mass and length.
However, the maximal  non-commutativity of  momenta and coordinates is
accomplished in the frame of the generalized Snyder-Yang algebra (GSYA)
with  three new constants with dimensionality of mass, action and length,
additional to the standard ones $c$ and $\hbar$ \cite{lezkh,lezkh1,
lezkh2,lezkh3}.

    For the considered below model of color particle confinement 
it is sufficient to use a special Snyder-Yang algebra (SSYA)  
for the generalized momenta and coordinates
 with some constraints on the new constants. 
We estimate  new constants'
values with the help of relativistic modification of
the constituent quark mass notion and  values of current quark masses
and energies of constituent quarks.

\smallskip
\begin{center}
\noindent {\bf 2. GENERALIZED SNYDER-YANG ALGEBRA}
\end{center} 
\smallskip

As is known  the QCD operates with 
 quantum color fields of quarks and gluons defined in the conventional
four dimensional Minkowski spacetime $M_{1,3}$ \cite{yndu}.
Generators of the Poincare group together with  generators
of translations in the $M_{1,3}$, {\it i.e.} the operators of 
momenta, coordinates and angular momenta, constitute a basis for an algebra
of observables of the conventional quantum theory. However,  in the general 
case  the coordinates and the momenta of a quantum particle
can be non-commutative between themselves and commutation relations
can depend on three new constants besides the Plank constant $\hbar$ and the 
velocity of light $c$. We consider the GSYA 
 in the following  form \cite{lezkh4}:

\[
[F_{ij}, F_{kl}]=i(g_{jk}F_{il}-g_{ik}F_{jl}+g_{il}F_{jk}-g_{jl}F_{ik}),
\]
\[
[F_{ij}, p_{k}]=i(g_{jk}p_{i} - g_{ik}p_j), 
\]
\[ [F_{ij}, q_k]=i(g_{jk}q_i - g_{ik}q_j),
\]
\begin{equation}
[F_{ij}, I]=0, \quad [p_i, q_j]=i(g_{ij}I + \kappa F_{ij}),
\label{al1}
\end{equation}
\[  
[p_i, I]=i(\mu^2q_i - \kappa p_i),   [q_i, I]=i(\kappa q_i-\lambda^2p_i),
\]
\[
[p_i, p_j ]=i\mu^2F_{ij},  \quad [q_i, q_j]=i\lambda^2F_{ij}, 
\]

\noindent where  $i, j, k, l = 0, 1, 2, 3$, $c = \hbar = 1$,  $F_{ij}$, $p_i$, 
$x_i$ are the generators of the Lorentz group and the operators of momentum 
components and coordinates, correspondingly, $I$ is an additional 
"identity operator"
($I$ is named as the "identity operator" in the quotation marks, 
because it goes to the  identity in the limiting 
case).  The new quantum constants $\mu$
and $\lambda$ have dimensionality of mass and lenght correspondingly.
The constant $\kappa$ is dimensionless in the natural system of units.

\smallskip
\begin{center}
\noindent {\bf 3. CONFINEMENT OF COLOR PARTICLES}
\end{center} 
\smallskip

By applying the algebra (\ref{al1}) to the description of color particles
the condition $\kappa = 0$ can be imposed. Actually it is known 
 nonzero $\kappa$ leads to the $CP-$violation \cite{lezkh,lezkh1,
lezkh2}, but  strong interactions   are invariant 
with respect to the $P-$, $C-$ and $T-$transformations on the high level 
of precision.  Moreover for color particles we imply the following relation:
$\mu\lambda = 1$. Thus for strong interaction color particles  we have the 
reduction of the GSYA to the special Snyder-Yang algebra (SSYA) in the case
that the constraints $\mu\lambda = 1$ and $ \kappa = 0$  are fulfilled. 
 Denoting $\mu$  as $\mu_c$ and  $\lambda$ as $\lambda_c$ we rewrite
 the  commutation relations (\ref{al1}) (without the standard commutation
relations of $p_i, x_i, F_{ij}$ with the Lorentz group generators
$F_{ij}$) as

\[
[p_i, q_j]=ig_{ij}I, [p_i, I]=i\mu_c^2q_i,  [q_i, I]=-i\lambda_c^2p_i,
\]
\begin{equation}
  [q_i, q_j]=i\lambda_c^2F_{ij},  \quad [p_i, p_j ]=i\mu_c^2F_{ij}.
\label{al2}
\end{equation}

    So we consider the generalized  model for a color particle motion,
when coordinates and momenta are on equal terms and form an eight dimensional  
 phase space: $h=$$\{h^{A}|h^{A}=q^{j},A=1,2,3,4,$ $ j =0,1,2,3,$
$h^{A}=$ $\tau p^{j},A=5,6,7,8,$ $ j =0,1,2,3\}$. 
$P=$ $\{P^{A}|P^{A}=$ $p^{j},A=1,2,3,4,$ 
$ j =$ $0,1,2,3,$ $P^{A}=$ $\sigma q^{j},A=$ $5,6,7,8,$ $ j =0,1,2,3\}$.
The constants $\tau$ and $\sigma$ have dimensions of length square and 
mass square, correspondingly. 
The generalized lenght square 
\begin{equation}
  l^2=h^Ah_A,
\end{equation}
\noindent and the generalized mass square 
\begin{equation}
  m^2=P^AP_A,
\end{equation}
\noindent are invariant under the O(2,6) transformations
in the phase space of a color particle \cite{lezkh4}, 
where $h_A=g_{AB}h^B$,
$g_{AB}=$$ g^{AB}=$$diag$$\{1,-1,-1,-1,1,-1,-1,-1\}$. 

Thus 
\[
dm^2 = (dp_0)^2 -(dp_1)^2 - (dp_2)^2 - (dp_3)^2
\]
\[
+ \sigma^2(dq_0)^2-\sigma^2(dq_1)^2 -\sigma^2(dq_2)^2 -\sigma^2(dq_3)^2 
\]
\begin{equation}
= (dp)^2+\sigma^2(dq)^2,
\end{equation}
\noindent 
and the coordinates $q^{i}$ and the momentum  
components $p^{i}$ are the quantum operators satisfied eqs.(\ref{al1})
 or eqs.(\ref{al2}) in the frame of this approach.

Under these conditions a new Dirac type equation for a spinorial
field $\psi$ has the following form:
\begin{equation}
  \gamma^AP_A\psi = m\psi,
\label{newd}
\end{equation}
\noindent where 
$\gamma^A$ are the Clifford numbers for the 
spinorial O(2,6) representation, i.e.
\begin{equation}
  \gamma^A\gamma^B+\gamma^B\gamma^A = 2g^{AB}.
\end{equation}

One can take the product of Eq.(\ref{newd}) with 
$\gamma^AP_A+m$ and apply Eqs.(\ref{al1}), then the following 
equation for $\psi$ can be obtained
\[
(p^ip_i + \sigma^2q^iq_i + 2\Sigma_{i<j}S^{ij}F_{ij} +
\]
\begin{equation}
  + 2\sigma S^0I)\psi = m^2\psi,  
\label{newds}
\end{equation}
\noindent where an explicit form of the coefficients $S^{ij}$ have been 
written in Ref.\cite{lezkh4}.

Eq.(\ref{newds}) contains the oscillator potential, which restricts
a motion of a color quark and causes  its confinement.  Note that  
Eq.(\ref{newds}) can also be
applied for a description of a confinement of boson particles such as 
diquarks and gluons with the same confinement parameter $\sigma$.

\smallskip
\begin{center}
\noindent {\bf 4. ESTIMATIONS OF QUARK PARAMETERS AND THE $\mu_c$ AND 
$\lambda_c$ CONSTANTS}
\end{center} 
\smallskip

Let us consider some consequences of this approach for 
determination of color quark characteristics.
   From the relations (\ref{al2}) it immediately follows nonzero
uncertainties for results of simultaneous measurements of quark
momentum components. For instance, let $\psi_{1/2}$ is 
a quark state with a definite value of its spin component along the third axis.
Consequently, 
\begin{equation}
[p_1, p_2] = i\mu_c^2/2, 
\end{equation}
thus
\begin{equation}
\Delta p_1 \Delta p_2 \ge \mu_c^2/4
\end{equation}
     and if $\Delta p_1 \sim \Delta p_2$, one gets 
\begin{equation}
\Delta p_1 >\mu_c/2, \quad
\Delta p_2 >\mu_c/2. 
\end{equation}
 We see that the generalized quark momentum components $p_{\perp1,2}$ 
cannot be 
measured better than tentatively one-half a  value of $\mu_c$ \cite{lezkh4}.

One can get  estimations of the $\mu_c$ and  $\lambda_c$ values
  using the quark equation (\ref{newds}). As it is seen,  
$m^2$ and $p^2$ entered into  Eq.(\ref{newds})
can be considered as current and constituent  quark masses squared, 
respectively.  So Eq.(\ref{newds}) indicates that the convential 
relation for a current quark $p_{cur}^2=m^2 $
 should be transform to 
\begin{equation}
p^2=k_*^2+ M^2, \quad  M=m+\Delta
\label{mcon}
\end{equation}
for a constituent quark, where $M$
is a constituent mass,  $k_*$ is an effective
value of the quark momentum. $\Delta$ and $k_*$ take into account the 
contributions from
the additional terms of Eq. (\ref{newds}).
To estimate the $\mu_c$ and  $\lambda_c$ values with the help of 
the constituent quark mass $M$ and the current quark mass $m$   
 a ground state $\psi_0$  in a meson has been considered  neglecting
 an orbital angular momentum contribution $L\psi_0$. Together with 
Eqs. (\ref{mcon})  we use the values of quark energies and its masses
evaluated in the framework of the relativistic model of quasi-independent
quarks \cite{lezkh5}.
By this means we obtain the following parameters of the 
constituent and current $u-$, $d-$, $s-$, $c-$, $b-$ quarks
within a few percents of the relative uncertainty. 
\begin{center}
{\it Table 1.} {\small Parameters of the constituent (Q) and the current (q) 
quarks in {\it MeV}'s at $k_*=130 MeV$}

\smallskip

\begin{tabular}{c|c|c|c|c|c}
\hline
{\it Quark}&u&d&s&c&b\\
\hline
{\small Q energy} &335&339&486&1608&4950\\
\hline
{\small Q mass} &309&313&468&1603&4948\\
\hline
{\small q mass} &5&9&164&1299&4644\\
\hline
\end{tabular}
\end{center}
The quark parameters written in the Tabl. 1 are not in contradiction
with the values of these characteristics, which have been determined
as in the frameworks of various models, as in the QCD frameworks
\cite{borka, pdg}.
Notice that the parameters presented above are evaluated
at the  130 MeV scale. Now that we have obtained the quark parameters
and $k_*$, the $\mu_c$ and $\lambda_c$ values are readily evaluated:
$\mu_c \approx 180  MeV$, $\lambda_c \approx 1.1 Fm$. 

\smallskip
\begin{center}
\noindent {\bf 5. CONCLUSIONS}
\end{center}
\smallskip

The convenience of the  considered  model of a color particle confinement
resides in its ability to include the dimensional parameters, namely
$\mu_c$ and $\lambda_c$, in  the relativistically invariant way
and in an explicit form. This  model brings into existence the
confining  particle, since in the framework of the model the  equations
of motion  (\ref{newd}) and (\ref{newds}) contain the  rising potentials 
which provides the confinement of the particles. 

It is interesting that $\mu_c$
value estimated above is approximatelly equal to the critical
temperature $T_c$ of the deconfinement phase transition.
If it is not accidental, then an adequate description of the quark-gluon 
plasma is impossible without incorporating
 a non-commutativity of dynamical observables in a like
manner as it is presented  in the Snyder-Yang algebra.

\end{document}